\documentclass[%
 aip,
 amsmath,amssymb,
 reprint,%
floatfix]{revtex4-1}

\usepackage{graphicx}
\usepackage{dcolumn}
\usepackage{color}
\usepackage{bm}

\usepackage[utf8]{inputenc}
\usepackage[T1]{fontenc}
\usepackage{mathptmx}

\begin{document}

\preprint{AIP/123-QED}

\title[]{\color{blue}The Role of Collective Elasticity on Activated Structural Relaxation, Yielding and Steady State Flow in Hard Sphere Fluids and Colloidal Suspensions Under Strong Deformation}
\author{Ashesh Ghosh}
\affiliation{Department of Chemistry}
\affiliation{Materials Research Laboratory}
\author{Kenneth S. Schweizer}
\email{\color{blue}kschweiz@illinois.edu}
\affiliation{Department of Chemistry}
\affiliation{Materials Research Laboratory}
\affiliation{Department of Material Science \& Engineering}
\affiliation{Department of Chemical \& Biomolecular Engineering \\University of Illinois at Urbana-Champaign, Illinois 61801, USA}


\begin{abstract}
We theoretically study the effect of external deformation on activated structural relaxation and elementary aspects of the nonlinear mechanical response of glassy hard sphere fluids in the context of the nonequilibrium version of the Elastically Collective Nonlinear Langevin Equation (ECNLE) theory. ECNLE theory describes activated relaxation as a coupled local-nonlocal event involving local caging and longer range collective elasticity, with the latter becoming more important with increasing packing fraction. The central new question is how this physical picture, and the relative importance of local caging versus collective elasticity physics, depends on external stress, strain and shear rate. Theoretical predictions are presented for deformation induced enhancement of mobility, onset of relaxation speed up at remarkably low values of stress, strain or dimensionless shear rate, thinning of alpha time and viscosity with apparent power law exponents, a non-vanishing activation barrier in the shear thinning regime, a Herschel-Bulkley form of rate dependence of the steady state shear stress, exponential growth of dynamic yield stresses with packing fraction, and reduced dynamic fragility and heterogeneity under deformation. The results are contrasted with experiments and simulations and qualitative agreement is found. An overarching conclusion is that deformation strongly reduces the importance of longer range collective elastic effects for most, but not all, questions, with stress-dependent dynamic heterogeneity phenomena being qualitatively sensitive to collective elasticity. Overall, nonlinear rheology is a more local problem on the cage scale than quiescent structural relaxation, albeit with deformation-modified activated processes still important.
\end{abstract}

\maketitle
\section{Introduction}
Fundamental understanding of slow thermally activated dynamics and mechanics in deeply supercooled molecular, metallic and polymer liquids, and also dense colloidal and nanoparticle suspensions, under quiescent equilibrium conditions remains a major challenge in condensed matter physics, physical chemistry, and materials science \cite{p1,p2,p3,p4,p5,p6}. The response of deeply supercooled liquids and glasses to strong external deformation is even less understood, but is of enormous practical importance and scientific interest \cite{p3,p6,p7,p8}. However, the venerable phenomenological idea of Eyring \cite{p9} that relates deformation to stress driven lowering of activation barriers, and thus faster relaxation and flow, has been highly influential for interpreting experiments in diverse fields. Significant support for this basic concept has also been found in simulation, especially very recently \cite{p10}. 
\par
From a ‘first principles’, force-based, nonequilibrium statistical mechanics perspective, much work has been done based on ideal Mode Coupling Theory (MCT) \cite{p11,p12,p13,p14}. The central idea is shear advection reduces the dynamical caging constraints which are quantified by two-point static structural correlations. Of direct relevance to our work are the ‘Isotropically Sheared Hard Sphere Fluid’ model (ISHSM) and the corresponding parametrized schematic $F_{12}^{(\dot{\gamma})}$ model \cite{p15,p16,p17}. An ideal glassy state in MCT inevitably ‘melts’ under deformation due to destruction of correlated fluid structure with increasing shear. Phenomena such as shear thinning of viscosity, increasingly homogeneous dynamics under deformation, and yielding have been studied, and constitutive equations constructed. An arbitrary small but finite shear rate melts the ideal glass state based on the advected wavevector idea. Often numerical results for hard spheres are only available for a small packing fraction range of 0.51-0.52 that straddles the hypothetical MCT glass transition under quiescent conditions \cite{p18}. In contrast, most experiments and simulations probe a much broader range of significantly higher packing fractions, where under quiescent conditions activated processes are crucial for structural relaxation, a mechanism not taken into account in ideal MCT. At high enough shear rates MCT predicts a ‘perfect’ shear thinning exponent of -1 \cite{p15,p16,p17}, while experiments or simulations sometimes observe exponents as small as -0.80 that depend weakly on packing fraction \cite{p19,p20,p21,p22,p23,p24,p25,p26}. A related issue is the experimental and simulation observation that the onset of shear thinning occurs at remarkably small values of “renormalized” or “dressed” Peclet number (product of shear rate and equilibrium structural relaxation time) of $Pe\sim 0.01-0.001$ \cite{p20,p21,p22,p23,p24,p25}, decades smaller than the MCT prediction of $Pe\sim 1$ \cite{p12,p13,p14,p15,p16,p17}. This suggests dissipative relaxation mechanisms at ultra-low renormalized Pe are of crucial importance which are not captured by MCT.
\par
Concerning the question of “yielding”, MCT predicts the existence of a nonzero yield stress in the zero-shear rate limit below its ideal glass transition \cite{p12,p17}. The flow curve (steady state stress versus shear rate) showshas an empirical Herschel-Bulkley (HB) functional form with an apparent nonzero “yield stress” at zero deformation rate and power law growth of steady state stress with shear rate. Such behavior is often seen (or inferred) in experiments performed over a limited range of shear rates. MCT also predicts the yield stress ($\Sigma^{+}$) follows a square root in packing fraction ($\phi$) law, $\Sigma^{+}\propto \sqrt{\phi-\phi_c}$ ) very close to the ideal glass transition at $\phi_c\approx 0.515$ \cite{p12}. However, experimental studies typically find an exponential growth of a dynamically determined yield stress ($\Sigma^{+}\propto e^{\# \phi}$, $\#$ is a large number, e.g. $\sim 50-60$) for a much broader range of packing fraction variation, which we believe is again suggestive of the crucial role of activated dynamics \cite{p26}. 
\par
A different and widely developed particle and force level statistical mechanical approach to glassy dynamics adopts the MCT idea to quantify dynamical constraints based on structural pair correlations, but formulates the problem in terms of stochastic single particle trajectories and captures activated hopping – the Nonlinear Langevin Equation (NLE) theory \cite{p27,p28}. This approach has no ideal glass transition below random close packing (RCP) since activation barriers are finite, and the ideal MCT transition becomes a smooth crossover to an activated barrier hopping dynamical regime. The central new concept is a particle-displacement-dependent “dynamic free energy”, the gradient of which defines a caging force on tagged particles \cite{p28}. Thermal fluctuation driven hopping is the primary relaxation mechanism in typical ultra-high concentration colloidal suspensions. Under macroscopic deformation, an external force on each particle is introduced in the dynamic free energy in a microrheological or mechanical work spirit. Its primary effect is to reduce the barrier of the dynamic free energy and particle jump distance, thereby accelerating relaxation and flow \cite{p29}. This nonequilibrium version of NLE theory has been widely employed for polymer glasses, and to a lesser extent colloidal glasses and gels \cite{p30,p31,p32,p33,p34}. It appears to capture well deformation-induced phenomena as a consequence of stress-induced softening of the localizing nature of the dynamic free energy. 
\par
However, recently it has been argued that quiescent activated glassy dynamics, in colloidal systems and even more so in viscous thermal (molecular, polymer) liquids, is strongly influenced by longer range collective elasticity effects in the deeply supercooled regime, in addition to the local cage physics embedded in NLE theory \cite{p35,p36}. Thus, the new question arises as to what is the role of collective elasticity for supercooled liquids and glasses under strong deformation? This is the topic of the present article, studied in the context of the hard sphere fluid. We ask whether local cage and collective elastic contributions to the dynamic activation barrier respond differently to external stress or strain? Regardless of the answer, studying how the successful ideas of ECNLE theory under equilibrium conditions change under deformation will provide a new window on testing basic concepts in glass physics, and the validity of the prior NLE-based work \cite{p29,p30,p31,p32,p33,p34} for nonlinear rheology. 
\par
Here we do not construct detailed constitutive models for specific deformations, but rather adopt a minimalist and physically transparent nonlinear Maxwell model which we believe captures much of the basic physics. The basics of the NLE \cite{p27,p28} and ECNLE \cite{p35,p36} theories under quiescent and strong deformation conditions \cite{p29} are reviewed in section II. Sections III and IV present our new analysis of how the consequences of collective elasticity change under strong deformation, at both the dynamic free energy level and for observables of experimental interest. Section V presents an initial analysis of how deformation can modify one metric of dynamic heterogeneity--stretched exponential relaxation and a distribution of relaxation times. We qualitatively and semi-quantitatively contrast our results with experiments and simulations. Section VI presents a summary and future outlook. The Appendix contains a brief analysis of the validity of the concept of effective temperature or effective density.
\section{Theory}
We consider a fluid of monodisperse hard spheres (HS) of diameter $\sigma$ and packing fraction of $\phi=(\pi/6)\rho\sigma^3$, where $\rho$ is the number density. Equilibrium pair structure is computed using Ornstein-Zernike (OZ) integral equation theory with the Percus-Yevick (PY) closure \cite{p37}. The key input to the dynamical theory is the static structure factor in Fourier space, $S(k)=1+\rho h(k)=(1-\rho C(k))^{-1}$, where $h(k)$ is the Fourier transform of the non-random part of the direct correlation function, $h(r)=g(r)-1$ and $C(k)$ is the direct correlation function \cite{p37}. Below we recall relevant aspects of the dynamical theory discussed in great detail previously. 
\subsection{Background: Quiescent State Dynamical Theory}
The starting point is a Generalized Langevin Equation (GLE) for ensemble-averaged tagged particle dynamics \cite{p27,p28}. The crucial quantity is the force-force time correlation function, which is computed based on naïve (single particle) Mode Coupling Theory (NMCT) as \cite{p28,p38},

\begin{equation}
\begin{split}
    K(t)&=\langle\vec{f}_0(t).\vec{f}_0(0)\rangle\\&
    =\frac{\beta^{-2}}{3}\int \frac{d\vec{k}}{(2\pi)^3}\rho k^2 C(k)^2S(k)\Gamma_s(k,t)\Gamma_c(k,t)
\end{split}
\end{equation}
where $\beta=(k_BT)^{-1}$ is the inverse thermal energy, $\Gamma_s(k,t)=\langle e^{i\vec{k}.(\vec{r}(t)-\vec{r}(0))}\rangle$ and $\Gamma_c(k,t)=S(k,t)/S(k)$ are the (normalized to unity at t=0) single and collective dynamic structure factors or propagators, respectively. A long time kinetically arrested state is modeled as an Einstein glass via a dynamic localization length, $r_L$ with Gaussian form of arrested propagators (Debye-Waller factors) as, $\Gamma_s(k,t\rightarrow \infty)=\exp({\frac{-k^2r_L^2}{6}})$ and $\Gamma_c(k,t\rightarrow \infty)=\exp({\frac{-k^2r_L^2}{6S(k)}})$. A self-consistent localization equation follows from the condition, $\frac{1}{2}\langle\vec{f}_0(t\rightarrow \infty).\vec{f}_0(0)\rangle r_L^2=\frac{3}{2}k_BT$, yielding \cite{p27,p28,p38},
\begin{equation}
\begin{split}
   \frac{1}{r_L^2}
    =\frac{\rho}{18\pi^2}\int_0^{\infty} dk k^4 C(k)^2S(k)\times \exp\Big(-\frac{k^2r_L^2}{6}(1+S^{-1}(k))\Big)
\end{split}
\end{equation}
Localization emerges at $\phi_c=0.432$ using PY closure [27]. The elastic shear modulus is \cite{p29,p39},
\begin{equation}
    G^{\prime}=\frac{k_BT}{60\pi^2}\int_0^{\infty} dk \Big[k^2\frac{d\ln S(k)}{dk}\Big]^2\times \exp\Big(-\frac{k^2r_L^2}{3S(k)}\Big)
\end{equation}
An analytic micro-rheology like relation has been derived for the dynamic elastic shear modulus in an idealized localized state \cite{p38,p39}, $\beta G^{\prime}\sigma^3\propto\phi \Big(\frac{\sigma}{r_L}\Big)^2$, and verified to reproduce numerical calculations based on Eq(3) very well.
\par
To go beyond the ideal MCT transition (signals a dynamic crossover to activated hopping) the stochastic Nonlinear Langevin Equation (NLE) theory for single particle trajectories has been formulated based on the concept of a dynamic free energy \cite{p27,p28}. Figure 1 shows a conceptual sketch of the theoretical ideas. The angularly averaged scalar displacement of a particle, r(t), obeys in the overdamped limit the stochastic NLE \cite{p27,p28}:
\begin{equation}
    -\zeta_s\frac{dr(t)}{dt}-\frac{\partial F_{dyn}(r(t))}{\partial r(t)}+\xi(t)=0
\end{equation}
where $\xi(t)$ is a white noise random force associated with the short time Fickian diffusion process quantified by the friction constant $\zeta_s$ and the dynamic free energy is, 
\begin{equation}
    \begin{split}
        \beta F_{dyn}(r)=-3\ln\Big(\frac{r}{\sigma}\Big)-\rho\int\frac{d\vec{k}}{(2\pi)^3}\frac{C(k)^2S(k)}{(1+S^{-1}(k))}\times &\\ \exp\Big(-\frac{k^2r^2}{6}(1+S^{-1}(k))\Big)
    \end{split}
\end{equation}
Beyond $\phi>\phi_c=0.432$, the dynamic free energy has a minimum at $r_L$, and a barrier of height $F_B$ at $r=r_B$. A representative result with relevant length scales indicated is shown in Fig. 1. 
\begin{figure}
    \centering
    \includegraphics[scale=0.83]{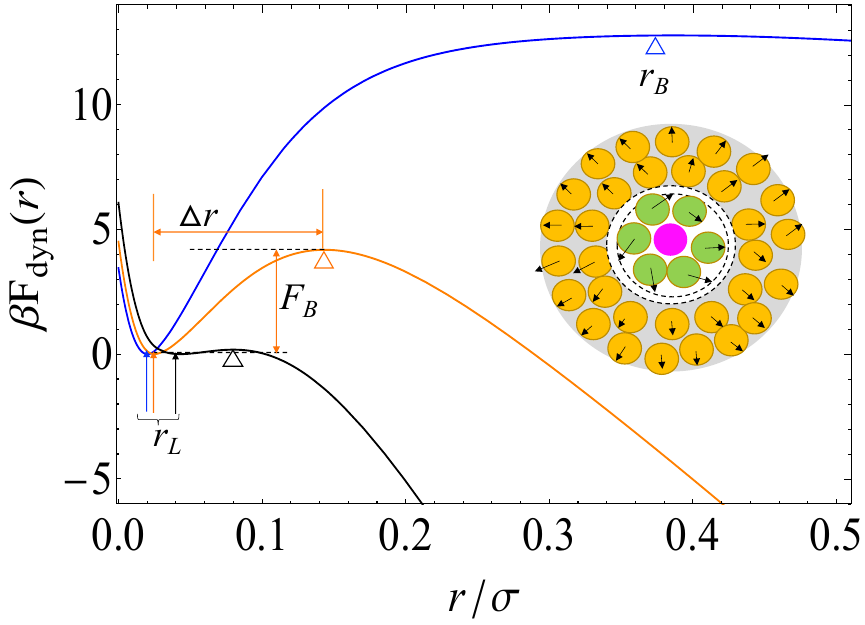}
    \caption{Dynamic free energy in thermal energy units as a function of particle displacement at a packing fraction of 0.61 for zero stress (blue), one-half the absolute yield stress (orange), and approaching the absolute yield stress (black), respectively. Important length scales are indicated. The cartoon shows the conceptual basis of ECNLE theory which involves a coupled local- nonlocal structural relaxation process. }
    \label{fig1}
\end{figure}
\par
NLE theory captures only local caging effects on single particle activated trajectories. It has been extended to include longer range collective effects in the Elastically Collective NLE (ECNLE) theory \cite{p35,p36}. The central idea, motivated by the phenomenological elastic “shoving model” \cite{p40,p41}, is all particles outside the cage must elastically displace via a spontaneous fluctuation in order to create the small amount of extra space required to allow large amplitude hopping of particles in the cage. This costs an elastic energy which contributes an additional barrier to the time scale for activated structural relaxation given by \cite{p36},
\begin{equation}
    \beta F_e=12\phi K_0\Big(\frac{r_{cage}}{\sigma}\Big)^3\Delta r_{eff}^2
\end{equation}
where $K_0=\frac{3k_BT}{r_L^2}$ is the harmonic spring constant of the dynamic free energy, $\Delta r_{eff}=\frac{3\Delta r^2}{32 r_{cage}}$ is the amplitude of elastic displacement field which at a scalar distance $r$ from the cage center is $u(r)=\Delta r_{eff}\Big(\frac{r_{cage}}{r}\Big)^2$ and $\Delta r=r_B-r_L$ is the particle jump distance. The alpha process is a coupled local-nonlocal event, and thus the total barrier is the sum of the local cage and collective elastic contributions, $\beta F_{Total}=\beta(F_B+F_e)$. The relaxation time follows from a Kramers mean first passage time analysis for barrier hopping as \cite{p36}
\begin{equation}
    \frac{\tau_\alpha}{\tau_s}=\frac{2}{\sigma^2}\times e^{\beta F_e}\int_{r_L}^{r_B} dx e^{\beta F_{dyn}(x)}\int_0^x dy e^{-\beta F_{dyn}(y)}
\end{equation}
where $\tau_s=\beta\zeta_s\sigma^2$ is a known short relaxation timescale, here taken as the elementary time unit. 
\par
A sample calculation of the alpha time is shown in Fig. 2. The elastic barrier is not important for lower packing fractions, but becomes increasingly dominant for $\phi\ge0.56$. The shear modulus is shown in the inset as discrete points, and is well described by an exponential law $\sim e^{25\phi}$ for $\phi\in{0.55,0.61}$. The analytic relation  $G^\prime\propto\phi/r_L^2$ is essentially exact.
\begin{figure}
    \centering
    \includegraphics[scale=0.76]{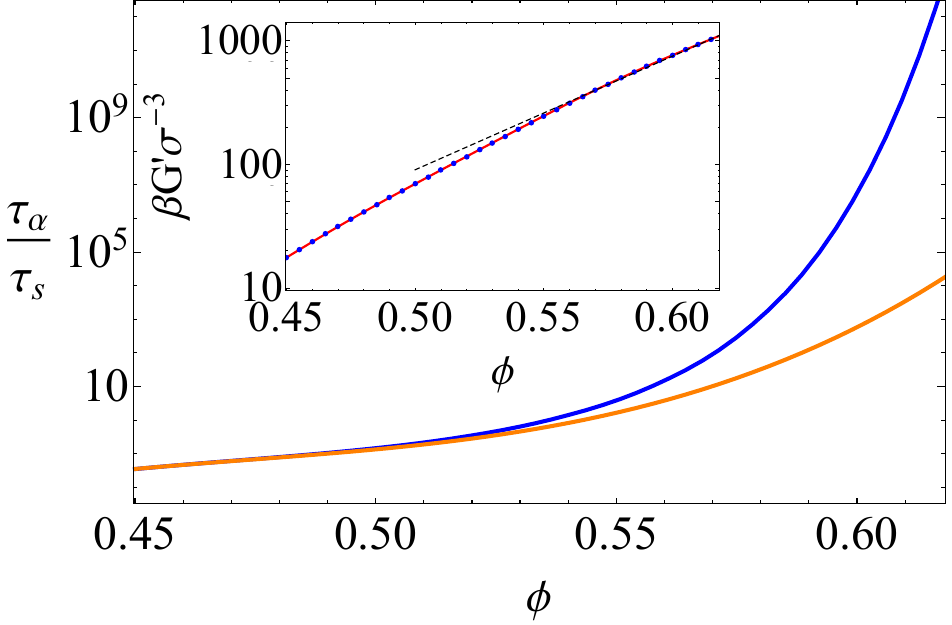}
    \caption{Non-dimensionalized alpha relaxation time of the quiescent hard sphere fluid as a function of packing fraction based on the NLE (orange) and ECNLE (blue) theories. Inset: log-linear plot of the dimensionless shear modulus as a function of packing fraction (blue points). The black dashed line shows a $\sim e^{25\phi}$ exponential dependence. The red solid curve is proportional to $\phi/r_L^2$and goes through all the theory data points.}
    \label{fig2}
\end{figure}
\subsection{Effect of External Forces and Stress }
Given the dynamic free energy controls single particle motion, to include external forces a stress-based micro-rheology perspective is adopted. Deformation enters the NLE evolution equation as a constant force, f, acting at the single particle level, corresponding to the non-equilibrium dynamic free energy acquiring a mechanical work contribution \cite{p29,p42,p43}:
\begin{equation}
    \beta F_{dyn}(r,\Sigma)=\beta F_{dyn}(r,\Sigma=0)-f.r
\end{equation}
where $\Sigma$ is stress and $f$ the magnitude of the external force. The connection of the latter two variables is a complex problem of force transduction, i.e. how macroscopic deformation is transmitted to the microscopic scale. The fundamental approximation is to assume a linear connection: $f=A\Sigma$, where $A$ is a microscopic cross-sectional area. Unique determination of the latter is not possible to within the uncertainty of a numerical prefactor. 
\par
Previous arguments which ignored the anisotropic nature of the deformation (consistent with the scalar description of particle trajectories) suggested \cite{p29} $f=\pi\sigma^2\phi^{-2/3}\Sigma/6$. Here we crudely estimate the mean consequence of anisotropy by noting an external force in a specific direction enters as $-fr\cos\theta$, where the angle between the applied force and particle displacement is $\theta$. Given NLE theory is formulated for particle trajectories at the level of a scalar displacement, a naïve pre-averaging of anisotropy is necessary and yields the factor $\frac{1}{4\pi}\int_0^{2\pi}d\phi\int_0^{\pi/2}d\theta \sin\theta\cos\theta=\frac{1}{4}$. The upper limit of the $\theta$ integration is $\pi/2$ since $\theta\in(\pi/2,\pi)$ models a compression zone where particles are moving in opposite direction of applied force and will be very slow. This analysis suggests:
\begin{equation}
    A=\pi\sigma^2/24
\end{equation}
where the $\phi^{-2/3}$ factor is ignored here since it is close to unity for the high packing fractions of interest. One could hypothesize $ A=\pi\sigma^2/4$ (particle cross sectional area) or other choices which only modify the numerical prefactor connecting stress and microscopic force. Below we take $A=\pi\sigma^2/24$ as the baseline choice, but provide a few examples and discuss more broadly how our central results change very weakly if an alternative extreme of $A=\pi\sigma^2/4$ is adopted. \par
An additional minimalist simplification is external forces are assumed to not change $S(k)$ on the dynamically-relevant cage scale \cite{p29}. There is simulation evidence this is a reasonable simplification \cite{p44,p45}, and this simplification is consistent with the scalar description of particle displacements. Finally, the short time relaxation process is even more local and is characterized here by a deformation-independent elementary time scale, $\tau_s$.\par
Figure 1 shows an example of how the dynamic free energy evolves with applied stress. The localization length increases, the jump distance and barrier decrease, and ultimately the localized form of the dynamic free energy is destroyed. The latter occurs at a critical force (equal to the maximum restoring force of the quiescent dynamic free energy), and the corresponding stress is defined as the “absolute yield” stress, $\Sigma_y^{abs}$. The latter is a theoretical construct which is an upper bound on the dynamic yield stress for Brownian systems \cite{p29,p44} which corresponds to when the deformation accelerated activated relaxation time becomes commensurate with the experimental time scale in the spirit of a stress-induced solid-to-liquid transition. This point also connects naively to MCT which ignores hopping and hence the absolute yield stress signals the destruction of the ideal MCT glass transition. It also seems germane to a “granular” scenario where thermal fluctuation driven uphill barrier hopping is impossible, and flow occurs only at a stress beyond that required to destroy the localization minimum of the dynamic free energy. Although the precise value of the absolute yield stress depends on A, the dimensionless ratio $\Sigma/\Sigma_y^{abs}$ does not, and hence calculations in this format are insensitive to the prefactor in A.\par
Another general aspect of Fig. 1 is that upon increasing stress to close to $\Sigma_y^{abs}$, although the barrier is destroyed, the localization length increases only modestly compared to the large decrease of the barrier location and jump distance.  As explained below, this is the key origin of the very different responses of the local cage versus collective elastic barrier to applied force. Finally, the stress dependent relaxation time follows from the Kramers’ mean first passage time expression but where all dynamic free energy quantities are now stress-dependent \cite{p29,p43}
\begin{equation}
    \frac{\tau_\alpha(\Sigma)}{\tau_s}=\frac{2}{\sigma^2}\times e^{\beta F_e(\Sigma)}\int_{r_L}^{r_B} dx e^{\beta F_{dyn}(x,\Sigma)}\int_0^x dy e^{-\beta F_{dyn}(y,\Sigma)}
\end{equation}
\subsection{Non-linear Maxwell Model}
This article does not construct a full rheological theory for any specific deformation type, but rather adopts a simple nonlinear Maxwell type of description based on how stress modifies the elastic shear modulus and mean alpha relaxation time. We believe such a minimalist description does capture much of the important physics, including strain softening of the elastic modulus, dynamic yielding in Deborah number space, and the steady state flow curve.  \par
The simple nonlinear elastic equation-of-state previously adopted (relevant in practice at times short compared to the structural relaxation time) implicitly defines strain as \cite{p29,p43},
\begin{equation}
    \Sigma=\gamma\times G^{\prime}(\Sigma)
\end{equation}
Though perhaps intuitive, this time local relation is not a rigorous constitutive relation. It is most directly related to a nonlinear instantaneous step strain experiment at t=0+. An ‘‘absolute yield strain’’ of qualitative value can be defined as \cite{p29},
\begin{equation}
    \gamma_y^{abs}=\Sigma_y^{abs}/G^{\prime}(\Sigma_y^{abs})
\end{equation}
The prefactor in the cross-sectional area A just scales the stress and strain computed from Eq(11). We also consider an alternative model based on defining a nonlinear shear modulus as, $G^{\prime}=\frac{d\Sigma}{d\gamma}$\cite{p31,p46}. Integration of this relation yields,
\begin{equation}
    \gamma=\int_0^\Sigma \frac{d\Sigma^{\prime}}{G^{\prime}(\Sigma^{\prime})}
\end{equation}
which differs from Eq(11), though with little consequence as shown below. These relations can be evaluated using Eq(3) and how the localization length evolves with stress. Since stress suppresses localization, this shear modulus will decrease monotonically with stress or strain.\par
To crudely address dynamic yielding in frequency ($\omega$) space, we again adopt a nonlinear Maxwell model and write the nonlinear storage and loss moduli as \cite{p43},
\begin{subequations}
\begin{align}
G^{\prime}(\Sigma,\omega) &=G^{\prime}(\Sigma)\times \frac{(\omega\tau_\alpha(\Sigma))^2}{1+(\omega\tau_\alpha(\Sigma))^2}\\
G^{\prime\prime}(\Sigma,\omega) &=G^{\prime}(\Sigma)\times \frac{\omega\tau_\alpha(\Sigma)}{1+(\omega\tau_\alpha(\Sigma))^2}
\end{align}
\end{subequations}
This model is crudely germane to Large Amplitude Oscillatory Shear (LAOS) measurements at a fixed frequency that measure the first harmonic to extract a characteristic stress or strain at which the system undergoes “dynamic yield” at that applied frequency. The latter is typically defined as when $G^{\prime}=G^{\prime\prime}$ , corresponding to a critical stress Deborah number of unity criterion: 
\begin{equation}
    De=\omega\tau_\alpha(\Sigma_c)=1
\end{equation}
where $\Sigma<\Sigma_y^{abs}$ is always satisfied.\par
Finally, the steady state stress, flow curve, and viscosity obey \cite{p32,p46}
\begin{equation}
    \Sigma_{ss}(\dot{\gamma})=\dot{\gamma}\eta_{ss}(\Sigma_{ss})=\dot{\gamma}G^{\prime}(\Sigma_{ss})\tau_\alpha(\Sigma_{ss})
\end{equation}
where $\dot{\gamma}$ is the shear rate and $\eta_{ss}$ the steady state viscosity. This is a nonlinear self-consistent equation for the steady state stress. A high shear rate limiting value due to short time processes could be included, but is irrelevant for our present purposes. Note that $\Sigma<\Sigma_y^{abs}$ is assured. The steady state viscosity and alpha time are $\eta_{ss}=\Sigma_{ss}(\dot{\gamma})/\dot{\gamma}$ and $\tau_{ss}=\Sigma_{ss}(\dot{\gamma})/\dot{\gamma}G^{\prime}(\Sigma_{ss})$, respectively. Unless stated otherwise, dimensional stress is presented in units of $k_B T/\sigma^3$.
\section{Foundational Results}
\subsection{Elastic Modulus}
NMCT calculations of the shear modulus are shown in Fig. 3 as a function of stress in a doubly normalized (by quiescent modulus and absolute yield stress) manner. For different packing fractions a rough collapse is predicted. The modulus reduction begins at a very low stress or strain, and is modest in magnitude and gently decreasing. The curve crossings reflect the double normalized form of the plots and the lack of perfect universality in this representation. \par
\begin{figure}
    \centering
    \includegraphics[scale=0.76]{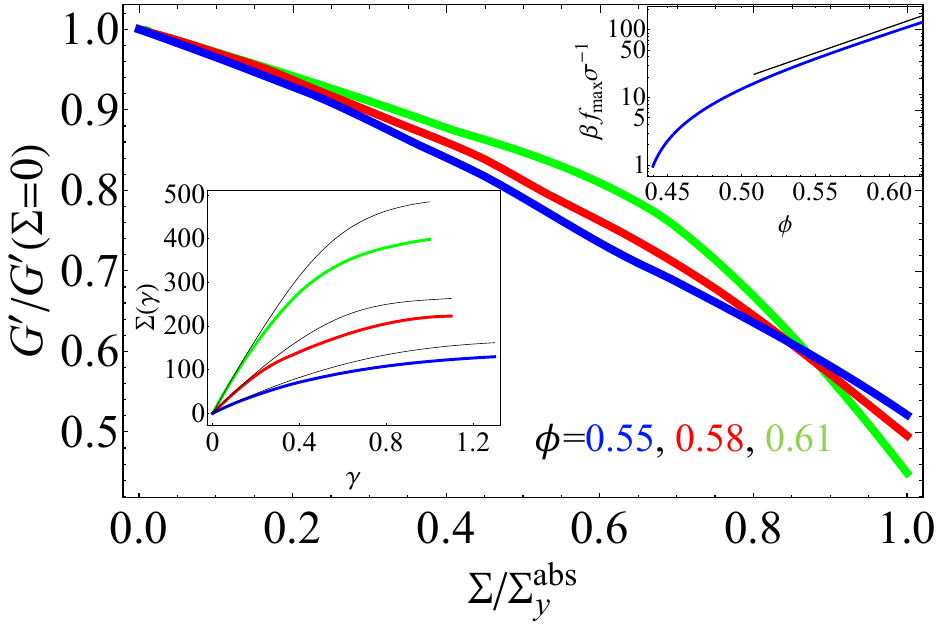}
    \caption{Naïve MCT calculation of the ratio of elastic modulus to its quiescent value as a function of normalized stress. Stress-strain curves based on Eq(11) are shown in the lower inset. Blue, red and green curves are for three packing fractions of 0.55, 0.58 and 0.61, respectively. The thin black solid curves in the inset are the analogous results using Eq(13). The applied force that destroys the localized form of the dynamic free energy is plotted as a function of packing fraction in the upper inset; the black line is an exponential dependence with an exponent of $\sim20$.}
    \label{fig3}
\end{figure}
The upper inset of Fig.3 plots in a log-linear manner the maximum cage restoring force of the quiescent dynamic free energy as a function of packing fraction. It equals the applied force required to destroy particle localization. At high densities a nearly exponential growth with packing fraction is predicted \cite{p28}, $e^{b\phi}$ with $b\sim 20$ for $\phi\in(0.53,0.61)$. The lower inset shows the stress-strain curves of the two nonlinear elastic models of Eqs (11) and (13), which produce nearly identical behavior. They terminate by construction at the absolute yield stress.  If the 1/24   prefactor of the cross-sectional area employed in Fig.3 is changed to 1/4, then we find nothing changes except the two axes of the lower inset are scaled by a factor of 1/6.
\subsection{Properties of Dynamic Free Energy}
\begin{figure}
    \centering
    \includegraphics[scale=0.76]{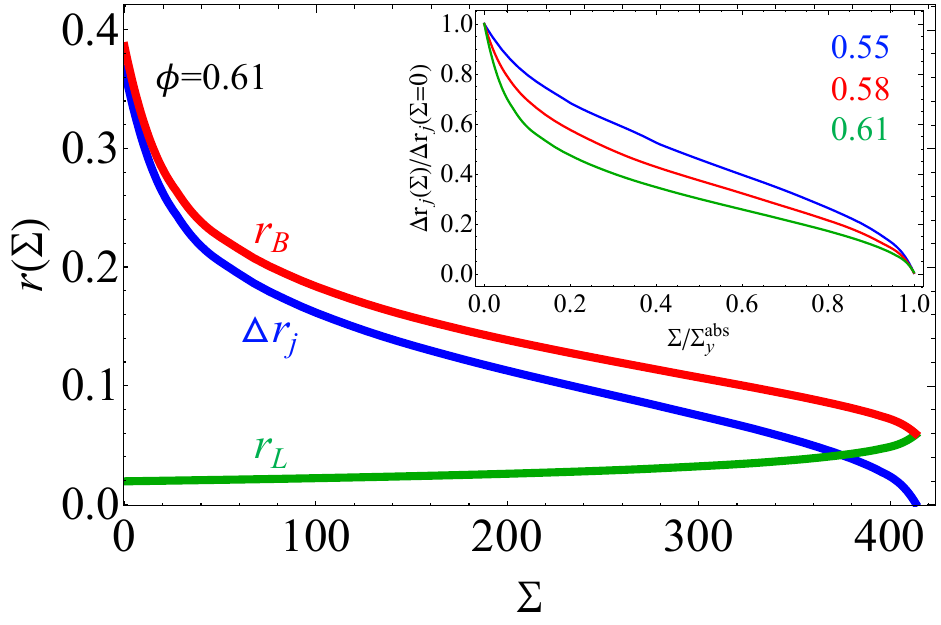}
    \caption{Localization length (green), barrier location (red), and jump distance (blue) in units of the hard sphere diameter as a function of stress for a packing fraction of 0.61. At absolute yield the barrier location and localization length merge. Inset: doubly normalized plot of the jump distance as a function of normalized stress for packing fractions of 0.55, 0.58 and 0.61 (top to bottom).}
    \label{fig4}
\end{figure}
An example of how the key length scales of the dynamic free energy change with stress is shown in Fig. 4. The localization length weakly grows, while the barrier location and jump distance strongly decrease. The inset shows the jump distance as a function of stress in a doubly normalized manner. In contrast to the localization length and elastic modulus, the jump distance shows a much worse collapse. In all subsequent plots and discussions, stress is the fundamental independent variable and strain is crudely estimated only for qualitative purposes using Eq(11). \par
Figure 5 shows an example of how the local and elastic components of the total barrier nonlinearly decrease with stress (or with strain in the inset). The key point is the elastic barrier vanishes much more quickly than its local cage analog. This seems intuitive since the elastic barrier reflects longer range physics, and its amplitude scales as the 4th power of the jump distance, per Eq(6). Hence, although collective elasticity is crucial for activated relaxation at high densities in equilibrium, its importance is relatively quickly rendered small and eventually negligible as stress or strain increases. This is a major new finding, and has a large impact on our results below, and the physical picture of linear versus nonlinear viscoelasticity, relaxation, and diffusion based on ECNLE theory. In detail, the “high deformation regime” begins when stress is only roughly one third of the absolute yield stress for all the high packing fractions studied. \par
\begin{figure}
    \centering
    \includegraphics[scale=0.76]{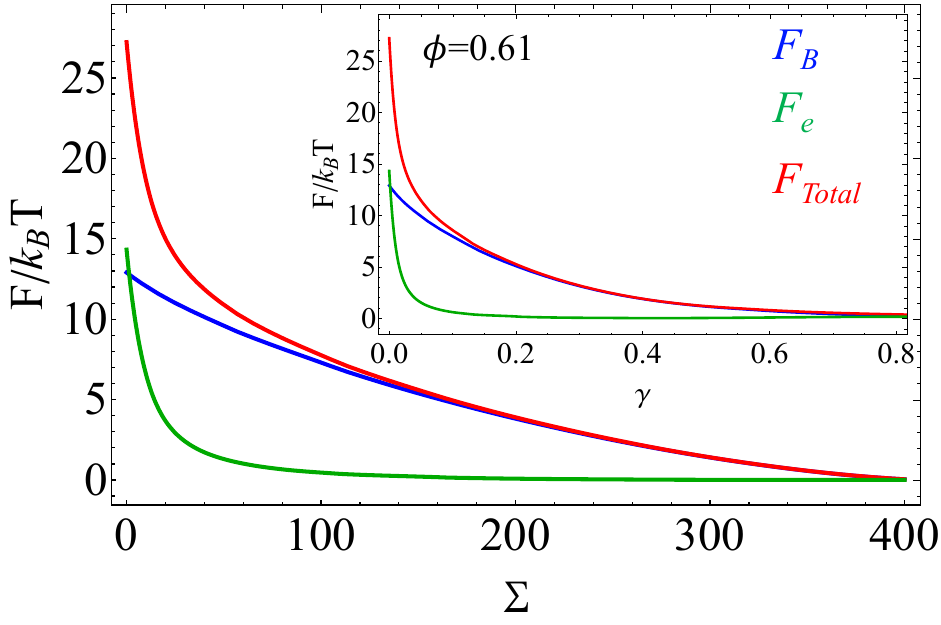}
    \caption{Local cage, collective elastic, and total barriers in units of thermal energy as a function of stress. Inset: same results as a function of strain $\gamma=\Sigma/G^{\prime}(\Sigma)$. The absolute yield stress (strain) is ~405 (~0.95) for the prefactor choice of $\pi/24$ in the cross-section area of Eq(9). }
    \label{fig5}
\end{figure}
The evolution of the local cage, collective, and total barriers with stress for 3 high packing fractions are plotted in a doubly normalized fashion in the main panel of Fig.6. An excellent collapse of the local barrier is found, but not the elastic barrier, and hence not the total barrier. The latter decays more rapidly at higher packing fractions since the elastic barrier is more important and is reduced much faster than its local analog with deformation. As an empirical curiosity, we find the theory predicts that the normalized elastic barrier curves are very well described (not shown) by their local analogs raised to an exponent $x(\phi)$ equaling $\sim$4.6, $\sim$6 and $\sim$8 for packing fractions of 0.55, 0.58 and 0.61, respectively. Such a connection between local and collective barriers is ubiquitous in ECNLE theory, and derives from the fact that all quantities needed to compute both barriers are determined by the same dynamic free energy. \par
\begin{figure}
    \centering
    \includegraphics[scale=0.76]{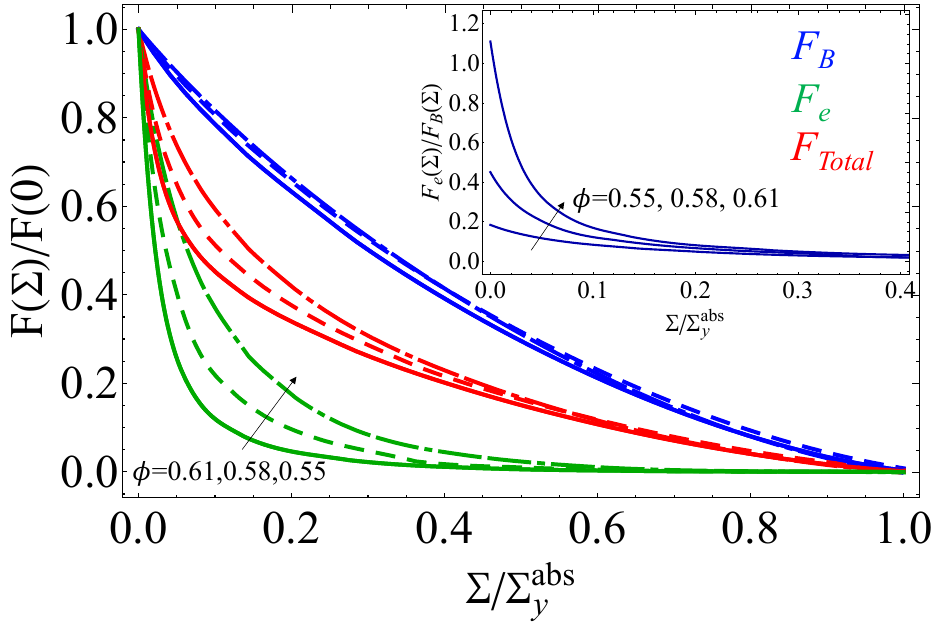}
    \caption{Doubly normalized local, elastic, and total barriers versus stress for packing fractions 0.61 (solid), 0.58(dashed), 0.55(dash dotted), respectively. The local barrier curves nearly collapse, but the elastic barrier curves do not. Inset: Ratio of the elastic to local barrier (measure of cooperativity and fragility in ECNLE theory) as a function of normalized stress. }
    \label{fig6}
\end{figure}
\par
The inset of Fig. 6 shows the ratio of the elastic to local barriers as a function of normalized stress. In equilibrium this ratio defines the ‘degree of cooperativity’ of the activated relaxation process and extent of non-Arrhenius behavior in ECNLE theory; it is closely related to dynamic fragility. The ratio decreases rapidly with stress, reinforcing the picture that deformation strongly and quickly reduces structural relaxation cooperativity. Hence, the prior studies based on NLE theory that successfully described the effect of active deformation on polymer glass relaxation and mechanics \cite{p29,p30,p31,p32,p46} will largely retain their validity.\par
\subsection{Activated Structural Relaxation Time}
Figure 7 shows representative results for the alpha time as a function of stress. The ECNLE and NLE theory calculations strongly differ in equilibrium (Newtonian plateau), and more so as packing fraction increases. But for all cases we find $\tau_\alpha^{ECNLE}\rightarrow\tau_\alpha^{NLE} $ at about one-third of the absolute yield stress, as expected from the inset of Fig.6.  A striking general feature of Fig.7 is that deformation begins to reduce the alpha time at very low values of reduced stress. The onset of this “stress thinning” behavior is at lower reduced stress when collective elasticity is included, and begins earlier at higher packing fractions where the elastic barrier is more important. A relatively good power law behavior is observed beyond the very low reduced stress of $\Sigma/\Sigma_y^{abs}\sim 0.01$ for the highest packing fraction ECNLE theory curve; an exponent of $\sim$ -5 captures $\sim$10 decades of alpha time thinning. The calculations in Fig.7 cover far more decades in time than what is measurable in simulation or colloid experiments. \par
\begin{figure}
    \centering
    \includegraphics[scale=0.76]{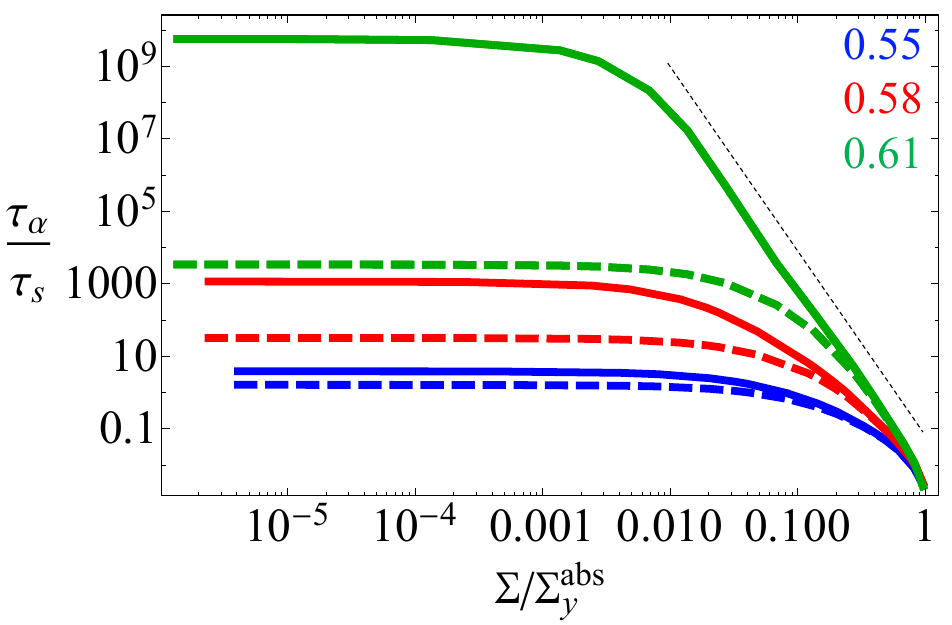}
    \caption{Double log plot of the nondimensionalized ECNLE and NLE theory alpha times as a function of normalized applied stress for packing fractions (top to bottom) of 0.61, 0.58 and 0.55. Note that $\tau_\alpha^{ECNLE}\rightarrow\tau_\alpha^{NLE}$ at a relatively low value of normalized stress. The black dashed line is a power law guide to the eye for the $\phi=0.61$ ECNLE theory curve. }
    \label{fig7}
\end{figure}
Plots of the relaxation time results in Fig.7 in a doubly normalized fashion reveals (not shown) a universal behavior for the NLE alpha time, per the excellent collapse found for local barriers in Fig.6. However, no such collapse is found for the ECNLE theory alpha time which involves the total barrier. Higher packing fraction alpha times decrease much more rapidly indicating the greater importance of collective elasticity. 
\par
\begin{figure*}
    \centering
    \includegraphics[scale=0.89]{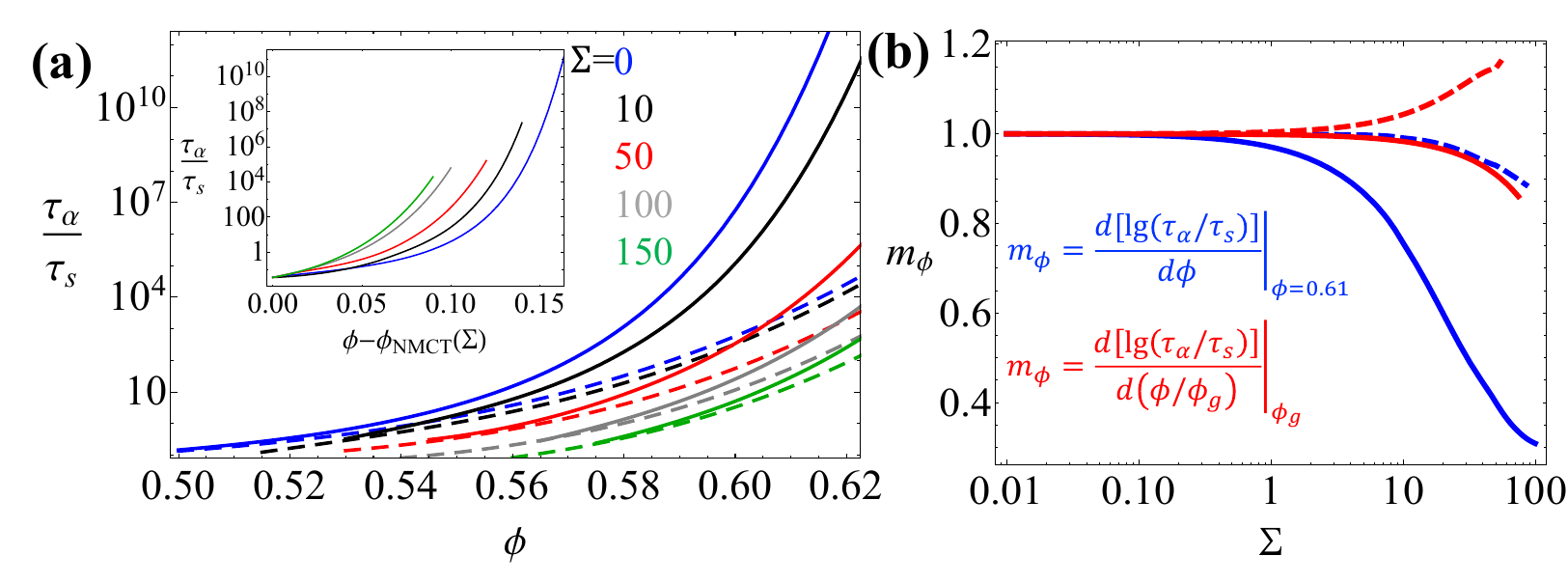}
    \caption{(a) ECNLE theory (solid) and NLE theory (dashed) dimensionless alpha times as a function of packing fraction for stresses of 0, 10, 50, 100, 150. As stress increases the slope of the ECNLE theory curves decrease a lot, but the NLE analogs do not. (b) Calculations of two definitions of fragility as a function of stress for the ECNLE (solid) and NLE (dashed) theories.}
    \label{fig8}
\end{figure*}
Figure 8a presents calculations of the alpha time versus packing fraction at various fixed stresses for ECNLE (solid) and NLE (dashed) theory. While the ECNLE curves show a strongly stress-dependent variation with packing fraction, the NLE theory analogs do not. With increasing stress, the NMCT crossover packing fraction that indicates the onset of activated dynamics grows. Roughly, we find its deviation from the quiescent state is,
\begin{equation}
\phi_{NMCT}(\Sigma)-\phi_{NMCT}(\Sigma=0)=a\sqrt{\Sigma}
\end{equation}
where $a\sim 0.008$. The inset of Fig.8a plots the alpha times as a function of “distance from the ideal localization transition”, $\phi-\phi_{NMCT}(\Sigma)$. It explicitly shows the alpha time grows more strongly at higher stresses at a “fixed distance” from the NMCT onset. The same conclusion is obtained for the NLE alpha times (not shown). We note that if alpha times are plotted as a function of $\phi/\phi_{NMCT}(\Sigma)$, similar qualitative conclusions are obtained (not shown). \par
We have examined how the glass transition packing fraction, defined in an objective kinetic manner from the temporal criterion of $\tau_\alpha(\phi_g(\Sigma))/\tau_s=10^x$, varies with stress. This relates to the problem of stress-induced fluidization of a glass. Calculations of $\phi_g^{ECNLE}(\Sigma)$ for a fixed stress were done for $x$=2,4,8,10,12 that covers up to 10 decades in timescale criterion. We find our numerical results are almost perfectly described by the simple linear form: 
\begin{equation}
    \frac{\phi_g^{ECNLE}(\Sigma)}{\phi_g(0)}\sim 1+a\Sigma
\end{equation}
where $a\sim 0.0012$ up to $\Sigma=20$. If the vitrification criterion corresponds to a much smaller time scale, more akin to simulations or colloid experiments, e.g., x=1,2,3,4, we find $\phi_g^{ECNLE}(\Sigma)<$0.64 up to $\Sigma=150$ and $\phi_g^{ECNLE}(\Sigma)/\phi_g(0)\sim 1+a\Sigma-b\Sigma^2$, where $b\sim a^2$. With increasing stress, the vitrification volume fraction difference between NLE and ECNLE theories decrease, again because of an earlier destruction of the elastic barrier with stress than the cage analog. \par
To further elucidate the role of collective elasticity under deformation a dynamic fragility is studied, defined in two ways. The first (i) definition is 
\begin{equation}
    m_\phi=\frac{d\Big(\log_{10}\frac{\tau_\alpha(\Sigma)}{\tau_s}\Big)}{d\phi}\Bigg|_{\phi_g=0.61}
\end{equation}
corresponding to the slope of the alpha time curve as a function of normalized packing fraction evaluated at a fixed $\phi_g=0.61$. The second definition (ii) employs a stress-dependent vitrification packing fraction: 
\begin{equation}
    m_\phi=\frac{d\Big(\log_{10}\frac{\tau_\alpha(\Sigma)}{\tau_s}\Big)}{d\Big(\frac{\phi}{\phi_g(\Sigma)}\Big)}\Bigg|_{\phi=\phi_g(\Sigma)}
\end{equation}
corresponding to the slope of the alpha time curve with normalized packing fraction that follows from the isochronal criterion, $\tau_\alpha(\phi_g,\Sigma)=\tau_\alpha(\phi_g=0.61,\Sigma=0)$. Results are shown in Fig. 8b. ECNLE theory predicts the fragility decreases with stress in both cases, and very strongly for definition (i). This seems intuitive, and qualitatively agrees with metallic glass simulations \cite{p47}, and some magnetic systems which show a decrease in fragility with applied magnetic field that tends to destroy the ‘spin-glass’ \cite{p48}. Although the NLE theory-based calculations yield a correct qualitative decrease in slope with deformation, the change in fragility from definition (ii) is the opposite of ECNLE theory and seems likely wrong. 
\section{Steady State Response }
\subsection{Yield Stresses and Shear Thinning}
We first consider the steady state flow stress described by the self-consistent Eq(16), and a dynamic yield stress defined from the Deborah number frequency domain criterion of Eq(15).  Calculations based on NLE and ECNLE theory are contrasted to reveal the role of collective elasticity. Recall again that the prefactor in A is expected to largely only modify the absolute stresses or strains, and not their dependence on packing fraction, frequency or shear rate.\par
The crossover “dynamic yield stress” determined from Eq(15) is plotted in Fig. 9 as a function of dimensionless frequency for two packing fractions. Differences between the ECNLE and NLE theory results are small. The ECNLE theory results converge to NLE results when elastic barriers are of small enough importance as realized at larger probing frequencies ($\tau_\alpha(\Sigma_c)=\omega^{-1}$). Figure 9 also shows the analogous steady state flow stress curves computed using Eq(16). The differences between the two theory results are visible, but small. Note that all “dynamic yield stresses” in Fig. 9 are below the absolute yield stress indicated by the arrows in Fig.9, as they must be since dynamic yielding in a stress-assisted dynamic barrier hopping event.\par
\begin{figure}
    \centering
    \includegraphics[scale=0.27]{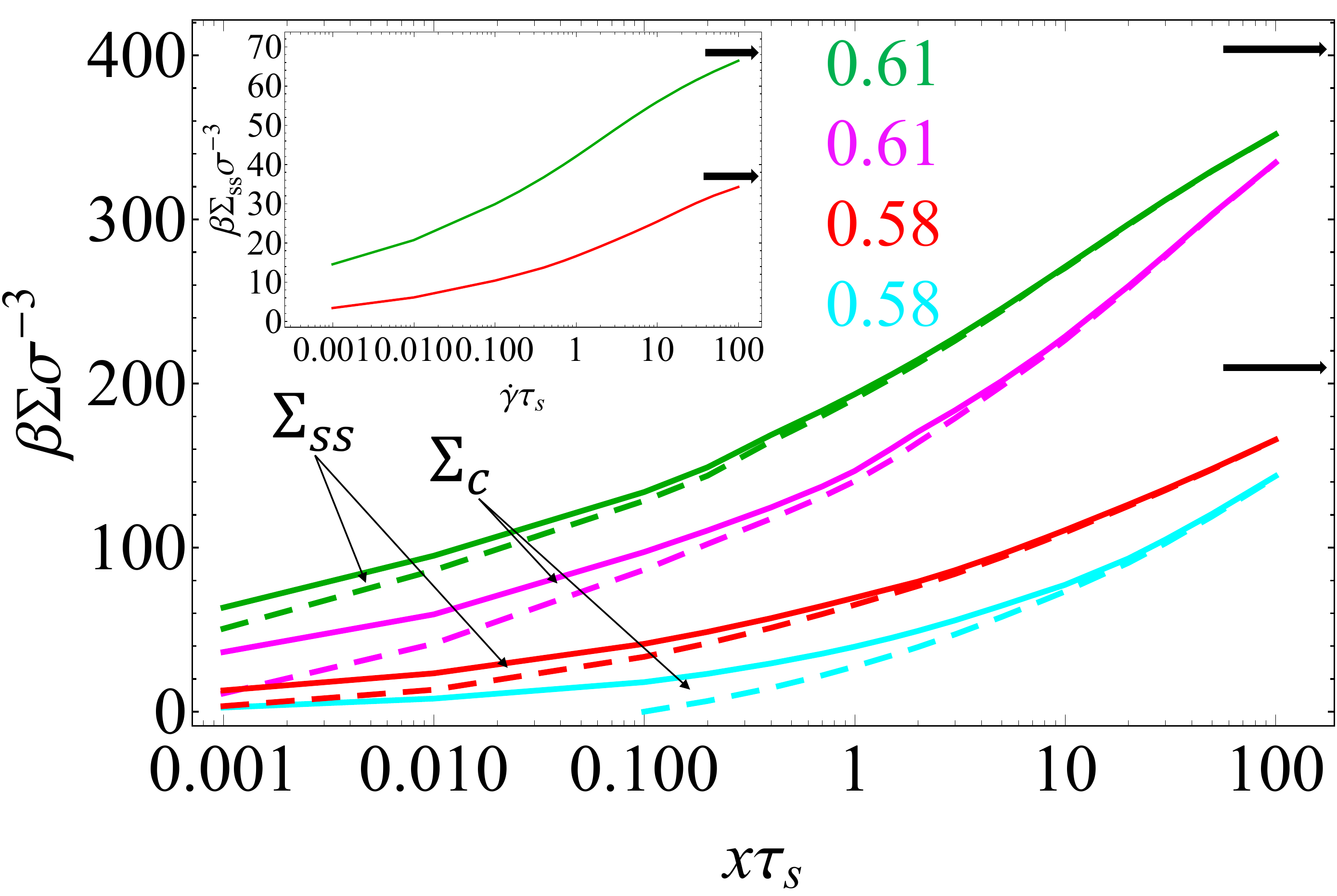}
    \caption{Critical crossover stress as a function of dimensionless frequency ($\omega\tau_s$), and the steady state flow stress as a function of dimensionless shear rate ($\dot{\gamma}\tau_s$), for packing fractions of 0.58 and 0.61. All stress curves can be well fit with the HB formula with exponents, n, of 0.18 and 0.27 for the crossover stress at packing fractions of 0.61 and 0.58, respectively, and 0.15 and 0.2 for the steady state stress at packing fractions of 0.61 and 0.58, respectively. Solid and dashed curves are based on using ECNLE and NLE theory, respectively. The corresponding absolute yield stresses are shown by black arrow on the right. Analogous steady state stresses based on $A=\pi\sigma^2/4$ are shown in the inset for packing fractions of 0.58 (red) and 0.61 (green).}
    \label{fig9}
\end{figure}
Recall that in both NLE and ECNLE theory there is always a finite structural and stress relaxation time in equilibrium, i.e., no Kauzmann transition \cite{p4} above RCP. Hence, at low enough frequency or shear rate, the system is always predicted to flow, and stress is linearly related to frequency or shear rate, and thus vanishes. But experimental and simulation data typically only cover 3-5 decades of shear rate or frequency \cite{p49,p50,p51}, and hence often cannot go to long enough times to check whether there is flow. Under this circumstance, data is often fit to the empirical Hershel-Buckley (HB) model expression of Eq(21) below. This motivates us to ask whether our theoretical curves in Fig. 9 can be described by the HB formula over the typical time scale range probed in practice. Recall the HB formula with exponent n is \cite{p51,p52},
\begin{equation}
    \Sigma(\dot{\gamma})=\Sigma_y^{HB}(\phi)+K(\phi)\dot{\gamma}^{n(\phi)}
\end{equation}
where $\Sigma_y^{HB}(\phi)$ is a packing fraction dependent “dynamic yield stress”, and $K(\phi)$ and $n(\phi)<1$ are known as the consistency index and flow index, respectively. \par
\begin{figure}
    \centering
    \includegraphics[scale=0.76]{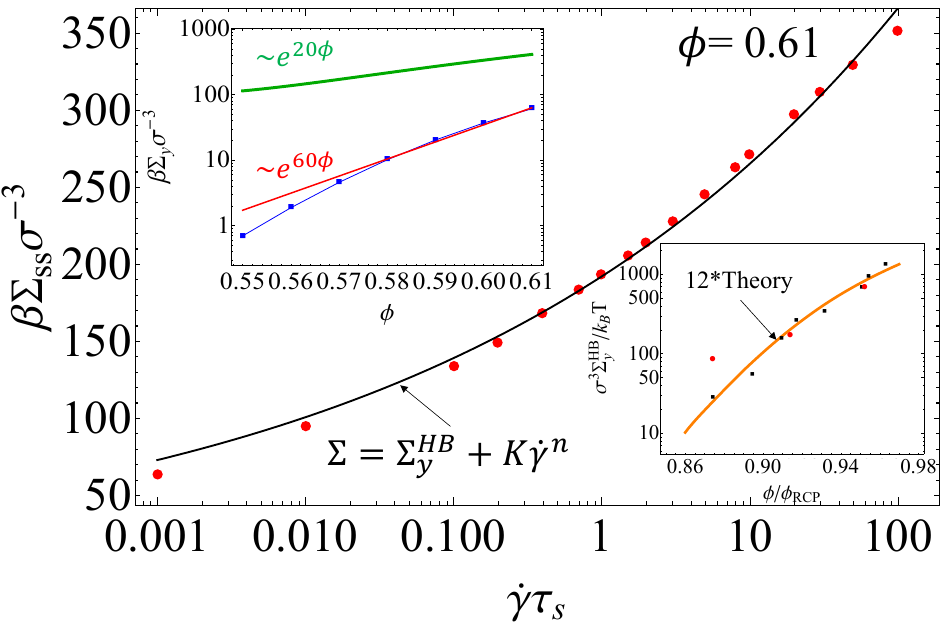}
    \caption{Log-linear plot of steady state stress (red points) for a packing fraction of 0.61. The black curve is a HB fit with an exponent of 0.15 and HB yield stress of $\sim$60. Top inset: yield stress extracted from fitting the theory results to the HB formula are plotted as a function of packing fraction. The red line is an exponential law. The absolute yield stress that fully destroys the dynamic free energy localized state is also plotted in green, and follows a much weaker exponential behavior. Bottom inset shows the experiment-theory comparison of the HB yield stress as a function of normalized packing fraction as discussed in the text. The theory curve has been shifted vertically by a factor of ~12, resulting in good agreement with experiment. }
    \label{fig10}
\end{figure}
As shown explicitly in Fig.10 for one example, we find that all our calculations in Fig.9 are well fit by the HB form, with an apparent exponent that decreases with packing fraction. For the $De(\Sigma)=1$ results, the exponents are 0.18 and 0.27 for packing fractions of 0.61 and 0.58, respectively; for the steady state flow stress results, we find exponents of 0.15 and 0.20 for packing fractions of 0.61 and 0.58. Regarding colloidal rheology experiments \cite{p17,p19,p26,p49,p50,p51}, the shear rates typically used are $\dot{\gamma}\sim 0.01-100 s^{-1}$, which for the ‘short-time Peclet number’ translates to $\dot{\gamma}\tau_s\sim 0.001-10$ based on $\tau_s\sim 0.1s$ for a $\sim100 nm$ sized colloid. Thus our calculations in Fig.9 that span $\dot{\gamma}\tau_s\sim 0.001-100$ are relevant to typical experiments. \par
The inset of Fig.9 shows analogous steady state flow curves based on the A prefactor in Eq(9) of 1/4 versus 1/24 (for the De number calculations the prefactor enters only as a trivial scaling factor and is irrelevant). Of course, the absolute stress values are smaller, and by a factor of $\sim$5. But the results are still well described by the HB formula, with slightly lower exponents of 0.13 and 0.17 for packing fractions of 0.61 and 0.58. \par
A representative example of the HB fits to our steady state stress predictions is shown in the main panel of Fig. 10 in a linear-log format; one sees good agreement over 5 decades in reduced shear rate. The limiting (and hypothetical) yield stress extracted from the HB fits are plotted as a function of packing fraction in the upper inset. The red line is $\sim e^{60\phi}$  and agrees well with our numerical results, especially for packing fractions in the range 0.58-0.61. Interesting, this dynamic yield stress shows a stronger packing fraction dependence than does the absolute yield stress, $\Sigma_y^{abs}\sim e^{a\phi}$ where $a\sim20$ for packing fractions of 0.55-0.61. \par
\begin{figure}
    \centering
    \includegraphics[scale=0.76]{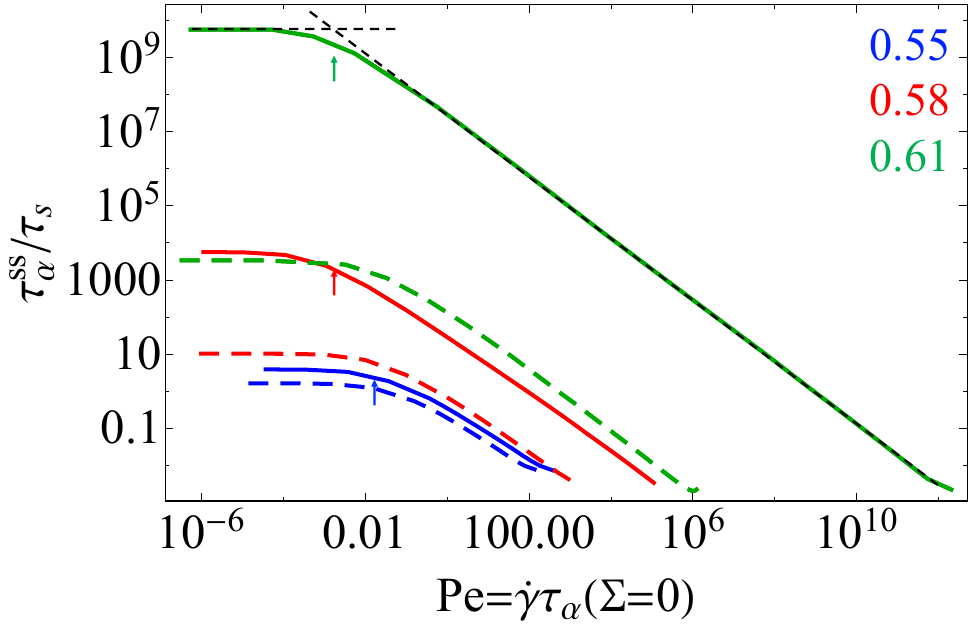}
    \caption{Non-dimensionalized (by the short time scale) steady state alpha time as a function of renormalized or dressed Peclet number for ECNLE (solid) and NLE (dashed) theories. The shear thinning regime can be described as an effective power law, and both theories predict very similar slopes at fixed packing fraction. A black dashed line of slope $\sim$ -0.80 is a fit for the packing fraction of 0.61 and describes the theoretical data over $\sim$10 decades of relaxation time. The colored arrows correspond to the crossover renormalized Pe number indicating the onset of shear thinning regime for the ECNLE alpha times.}
    \label{fig11}
\end{figure}
Figure 11 shows calculations of the steady state alpha relaxation time as a function of “renormalized” or “dressed” Peclet number ($\dot{\gamma}\tau_\alpha(\Sigma=0)$) for a wide range of packing fractions. The latter cover a range where the quiescent alpha times differ by $\sim$9 decades, per the plateau values in Fig. 11. The shear thinning region is empirically well fit by a power law of the form $\tau_\alpha^{ss}\propto \dot{\gamma}^{-\nu(\phi)}$ where $\nu(\phi)$ is a weakly packing fraction dependent exponent. The power law behavior works remarkably well for the ECNLE based calculation over ~9 decades of shear thinning for the highest packing fraction, as shown by the black dashed line. The effective exponents are 0.8, 0.77 and 0.75 for packing fractions of 0.61, 0.58 and 0.55, respectively, and nearly identical whether NLE \cite{p29} or ECNLE theory is employed.\par
Perhaps the most interesting prediction in Fig.11 is the onset of shear thinning occurs at very small values of renormalized $Pe <<1$. Naïve arguments, which work well in materials such as entangled polymer liquids where dynamics is not activated, is shear thinning is expected to begin when the renormalized Pe~1. We estimate a critical Peclet number, Pet, from the intersection of the Newtonian plateau and power law regime, and find $Pe_t=0.004$ (ECNLE, 0.61) and $Pe_t=0.03$ (NLE, 0.61). These values are enormously smaller than the naïve argument, and much smaller than predicted by MCT (of order unity) \cite{p12,p13,p14,p15,p16,p17}. Such huge differences relate to the fact that the physical process involves stress-assisted activated relaxation in our approach where deformation lowers the barrier. We also note that, although the NLE and ECNLE quiescent alpha times (the plateau values) can differ by nearly 6 decades, the $Pe_t$ values differ only by $\sim$1 order of magnitude (or less). Thus, the use of NLE theory in prior work is justified at zero order \cite{p29,p30,p31}. Of course, given the vastly superior nature of ECNLE theory for equilibrium relaxation compared to NLE theory, its predictions for deformation induced effects are more reliable. Similar plots for the steady state viscosity can be made as a function of dimensionless renormalized Peclet number (not shown). All the results are, as expected, very similar to that found for steady state alpha time.\par
Finally, we ask the physical question: what is the nature of particle trajectories at dynamic yield? Are they still intermittent and activated despite the enormous acceleration of relaxation by deformation? Prior combined confocal microscopy and mechanical experiments found the answer to be yes, even at high shear rates \cite{p19}. In our theoretical scenario, this would mean non-vanishing barriers in the dynamic free energy persist even at yield.\par
\begin{figure}
    \centering
    \includegraphics[scale=0.76]{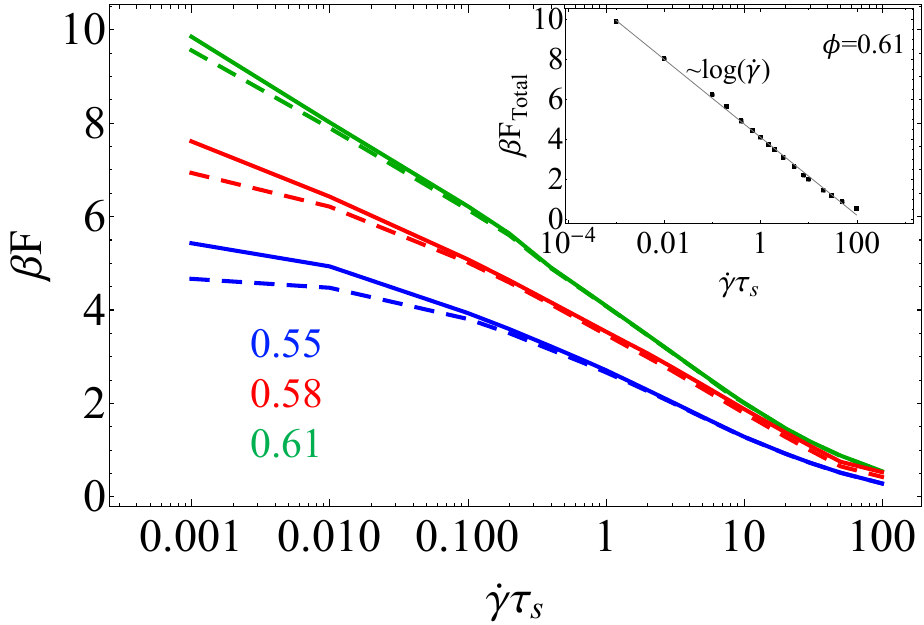}
    \caption{Local cage (dashed) and total (solid) barriers in the steady state as a function of dimensionless shear rate for packing fractions of 0.61, 0.58 and 0.55. The inset plots the ECNLE theory total barrier as a function of reduced rate for a packing fraction of 0.61. A logarithmic dependence is found over the range of shear rate range shown, in accordance with the power law shear thinning alpha relaxation time plot in Fig.11. We note that typical colloid experiments and simulations probe the limited range of short time Peclet numbers of 0.001 to 10.}
    \label{fig12}
\end{figure}
To address the above issue, in the main frame of Fig.12 we plot both the local cage and total barriers at dynamic yield. Collective elasticity effects are perturbative. But the barriers are generally not zero at yield, and are very substantial at lower effective rates relevant to practical experiments. Note the x-axis in Fig.12 scales as particle radius cubed at fixed packing fraction, and hence the use of smaller colloids would more incisively probe our prediction that yielded fluids still display activated trajectories. The x-axis also scales linearly with solvent viscosity which can be tuned. Overall, our results are consistent with the intermittent trajectory observations in experiment \cite{p19}. However, at ‘ultra-high’ shear rates (amenable to computer simulations) the barrier is predicted to approach zero at yield. This crudely connects our approach with an ideal MCT scenario where dynamics is not activated, and perhaps granular materials where thermally driven barrier hopping is absent and yielding is driven purely mechanically. Finally, the inset of Fig.12 shows the total barrier at yield decreases in a logarithmic manner with shear rate, consistent with power law thinning of the alpha time.
\subsection{Comparison to Experiments and Simulations}
There are many different protocols for introducing deformation, we have not performed full rheological calculations relevant to specific forms of deformation, real colloidal systems are poly-disperse, etc. These considerations render a quantitative comparison of our results with experiment and simulation inappropriate. But we believe it is important to qualitatively and semi-quantitatively contrast our findings with experiments and simulations. Below we do this and consider the dynamic yield stress, onset of shear thinning, and shear thinning exponent.
\par
Dynamic yield stresses are often determined in experiment from HB fits to steady state stress data. Such results from hard sphere (HS) colloid experiments with two different diameters \cite{p26} are plotted in the lower inset of Fig. 10, along with our (vertically shifted) theoretical predictions. The experiments employed polydisperse hard spheres and high packing fractions (RCP reported to be $\sim$0.67), while the theoretical results are for the monodisperse hard sphere system. This ambiguity motivates our plot of the theoretical and experimental data in terms of normalized packing fraction using RCP values of 0.64 (theory) and 0.67 (experiment). Using a vertical fit factor of ~12, one sees the theory curve agrees very well with experiment. This includes the subtle trend of an exponential growth with reduced packing fraction which weakly bends over at the highest packing fractions approaching (but still well below) RCP.
\par
Related experiments by Koumakis et. al. \cite{p49} measured startup shear stress-strain curves for hard sphere colloids of diameter $\sim$530 nm with 6\% polydispersity. Their highest packing fraction studied maps roughly to our monodisperse hard sphere packing fraction of $\sim$0.58. The experiments find an initial linear increase of stress, which begins to become nonlinear at very low values of strain (beyond ~1\%). This seems qualitatively consistent with our calculations in the inset of Fig.3.
\par
Molecular Dynamics (MD) simulations of 2D repulsive disk supercooled liquids by Furukawa et. al. \cite{p20,p21} have found remarkably small values of the renormalized Peclet number (product of shear rate and equilibrium alpha time) of $\sim$0.001-0.01 for the onset of viscosity shear thinning onset over a wide temperature range where alpha time differs by $\sim$4 decades. They observe ~3 decades of power law shear thinning with an effective exponent of $\sim$-0.8. Although there is no unique way to map these 2D simulations to our 3D monodisperse hard sphere model, they seem consistent with our calculations in Fig. 11. We again emphasize that the remarkably low dressed Pe number for the onset of shear thinning is a consequence of our theory activated relaxation. The results in Fig.11 are also consistent with confocal measurements \cite{p19} of the single particle alpha time at high packing fractions which exhibits a thinning exponent of $\sim$-0.8. 
\par
Although our focus has been on dense hard sphere colloidal suspensions, ECNLE theory has been extensively applied to molecular and polymer thermal liquids based on a mapping to an effective hard sphere fluid where the glass transition temperature, $T_g$, occurs at a packing fraction of $\sim$0.6-0.61 \cite{p53,p54}. This allows us to qualitatively compare our present results to experiments on polymer glasses that measured the segmental alpha relaxation time under deformation \cite{p8,p22,p23,p24,p25}. These measurements indicate a shear thinning onset at a very low renormalized $Pe <0.01$. Power law shear thinning is observed with an apparent exponent that varies from -0.77 (for $T_g-15K$) to -0.85 (for $T_g-25K$), qualitatively consistent with our theoretical results of thinning exponents of $\sim$-0.75 to -0.80 as packing fraction is increased from 0.55 to 0.61. 
\section{Dynamic Heterogeneity Under Deformation}
\subsection{Theoretical Results}
The theory discussed above takes into account only the mean relaxation time and how it evolves under deformation, per a minimalist Maxwell model. In glass forming liquids, there is a distribution of relaxation times, an effect often ascribed to “dynamic heterogeneity”. Its origin even under equilibrium conditions is still debated. However, very recently ECNLE theory for quiescent glass forming liquids has been extended to approximately take into account this type of dynamic heterogeneity, and has addressed problems such as diffusion-relaxation decoupling and stretching exponents of time correlation functions in colloidal and supercooled molecular liquids \cite{p55}. The model is based on a real space domain picture (on the 3-4 particle diameter scale) where the intrinsic local fluctuation of mean density gives rise to a distribution of dynamic free energies, and hence a distribution of alpha times. The physical ansatz is that all barrier fluctuations arise from the fluctuation in jump distance associated with the elastic barrier, which sets the amplitude of the elastic displacement field outside the cage; for details see ref. \cite{p55}.
\par
Motivated by experimental observations that dynamics in colloidal suspensions, polymer glasses, and supercooled metallic liquids become more `homogeneous' (transition from highly stretched to nearly single exponential relaxation) with applied deformation, we employ the recently extended ECNLE theory \cite{p55} to briefly consider the effect of deformation on this question. The generic alpha relaxation function, $C(t)$, is an average over exponential Debye contributions weighted by the predicted distribution of barriers. The numerical results are reasonably well fit by a stretched (KWW) exponential form, $C(t)=\exp\Big(-(\frac{t}{\tau^\star})^{\beta_K}\Big)$. 
\begin{figure*}
    \centering
    \includegraphics[scale=0.89]{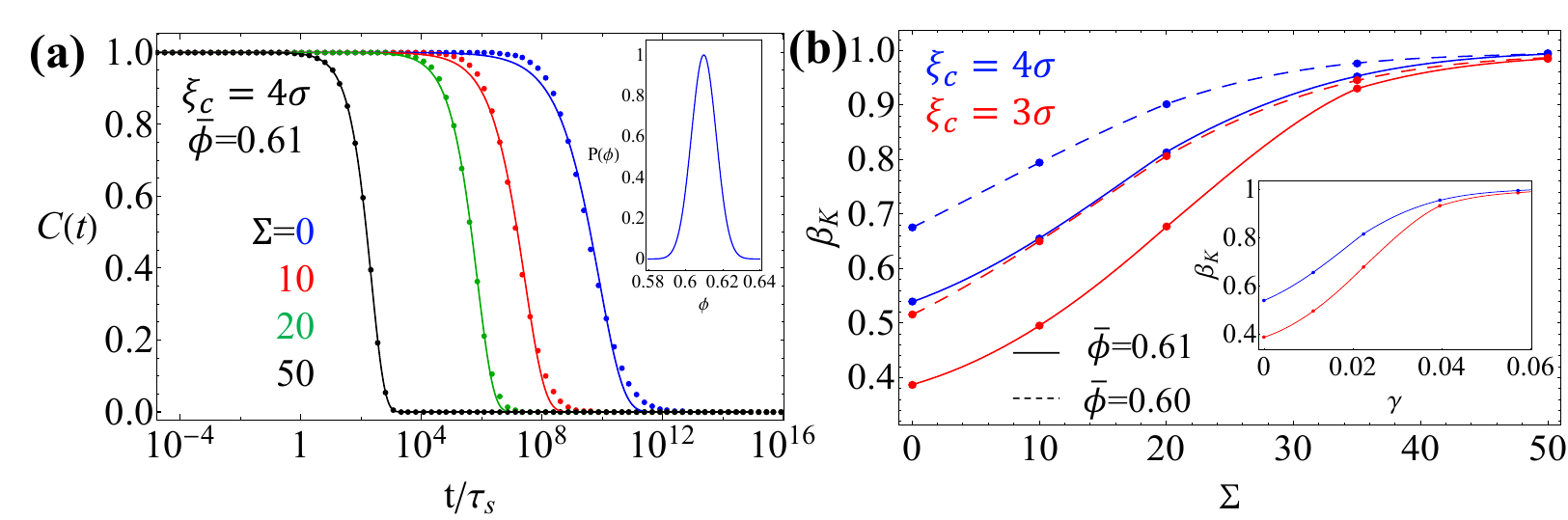}
    \caption{(a) Generic relaxation time correlation function at a fixed packing fraction of 0.61 and domain size of 4$\sigma$ at stresses of 0,10, 20 and 50. The solid curves are KWW fits of the discrete theoretical points. The inset shows the probability distribution function for local packing fraction fluctuations. (b) Extracted KWW stretching exponent for packing fractions of 0.61 (solid) and 0.60 (dashed) as a function of stress for two values of domain diameter. Inset shows the stretching exponent as a function of strain (using Eq(11))for a packing fraction of 0.61.}
    \label{fig13}
\end{figure*}
\par
Representative calculations of C(t) are shown in Fig. 13a for a single high packing fraction, different stresses, and domain diameter of 4$\sigma$. The KWW fits of the theoretical points are the solid curves. The extracted stretching exponents are plotted as a function of stress in Fig.13b for 2 packing fractions and 2 choices of domain size as motivated previously \cite{p55}. The stretching exponents in an equilibrium deeply supercooled state is $\sim$0.4-0.5, and then increases with stress, eventually approaching unity when the deformation effectively destroys the collective elastic barrier. Higher packing fraction and smaller domain size leads to more stretching, and a slightly delayed recovery of the dynamically homogenous limit where the stretching exponent approaches unity. 
\par
Importantly, the increase of the stretching exponent commences at low values of stress or strain (see inset), a direct consequence of the collective elastic origin of distribution of relaxation times in the deeply supercooled regime. Moreover, recovery of single exponential behavior is attained at low values of stress or strain. We remind the reader that the absolute stress and strain scales are sensitive to the numerical prefactor that enters the cross-section area A. If the latter is increased by a specified factor, then the x-axis in Fig.13b is roughly reduced by that factor. \par
Summarizing, by combining the recently developed theory for a distribution of alpha times \cite{p55} with the present analysis of deformation, we are led to a predictive model for how deformation reduces dynamic heterogeneity via its impact of stress on collective elasticity. 
\subsection{Comparison with Experiments and Simulations}
Molecular mobility experiments \cite{p8,p24,p25} of deformed polymer glasses below $T_g$ show significant dissipative processes emerge at very low (<<0.10) stress and strains. For the very low strain homogeneous deformation experiments, molecular mobility increased by $\sim$3 decades with the relaxation function becoming more ‘homogeneous’ with increasing deformation. The non-exponentiality of the correlation functions were fit to a stretched KWW form, and the exponent $\beta_K$ was found to grow from $\sim$0.3 to $\sim$0.8 with increasing deformation (strain) at $T_g$-20K. These results agree qualitatively with Fig.13b, and thus massive narrowing of the relaxation time distribution at modest deformations appears consistent with the ECNLE theory based analysis.\par
MD simulations of dynamical mechanical spectroscopy of model metallic glasses \cite{p47} have discovered a “fragile-to-strong” transition as mechanical strain is increased along with accelerated alpha relaxation and suppressed dynamic heterogeneity. Fragility normalized by its zero-strain value begins to decrease at very low strain amplitudes of 2\%, and drops by a factor of $\sim$5 for a 7.5\% amplitude strain. These simulations were done at relatively high temperatures where a glass transition is defined based on alpha times of order nanoseconds. The latter qualitatively map to the lower barrier regime of our theory with packing fractions $\sim$0.55-0.56. Our theoretical findings in Fig.8b based on a fixed time criterion suggest the same qualitative behavior per a fragile to strong transition based on using the ECNLE theory. \par
Finally, a few reports exist \cite{p56} suggesting (speculatively) that a causal relationship exists between stress or strain induced ‘glass melting’ and thermal fluctuation driven activated relaxation for non-stressed glasses. Specifically, the possibility of a mapping between a stressed glass and a higher temperature or lower density quiescent system has been advanced, in an “effective temperature” or “effective density” spirit. However, the applicability of such a connection remains strongly debated \cite{p47,p57}. We have briefly considered this issue in the Appendix, and find based on ECNLE theory such a connection is not supported.
\section{Discussion}
We have explored the role of deformation on the elastic shear modulus, structural relaxation time, viscosity, steady state flow stress, onset of shear thinning, apparent power law exponent of thinning, various metrics of yielding, and changes of a specific measure of dynamic heterogeneity using the microscopic ECNLE theory \cite{p35} in the context of dense hard sphere fluids and suspensions. Our analysis is based on a nonequilibrium formulation of the dynamic free energy concept \cite{p28}, and a minimalist nonlinear Maxwell model level of description. The latter corresponds to assuming the critical physical issues are how the elastic shear modulus and activated structural relaxation time (or equivalently the activation barrier) evolve under deformation. A prime goal is to understand how applied stress, strain and shear rate modify the fundamental local-nonlocal activated relaxation event associated with local caging and longer range collective elasticity in ECNLE theory.
\par
A primary finding is that relatively modest levels of stress or strain reduce the collective elastic barrier far more that its local cage analog. Well below any “yielding event”, the collective elastic barrier is essentially destroyed while its local cage analog, though also reduced by deformation, is still of significant magnitude. Hence, although quiescent structural relaxation under deeply supercooled or high packing fraction conditions is crucially impacted by longer range collective elastic effects beyond the cage scale, under highly nonlinear deformation conditions the problem is predicted to become more local and controlled for many questions mainly at the local cage scale. This finding has multiple important consequences. For example, it provides theoretical support for prior work that used NLE theory (no collective elasticity) to address the effects of deformation and nonlinear rheology in polymer glasses, and to a lesser extent particle glasses and gels \cite{p29,p30,p31,p32,p33,p34,p46}. Of course, collective elasticity still plays a role, and to varying degrees depending on the question. For example, it remains very important for understanding how deformation changes dynamic fragility and also the extremely low value of renormalized Peclet number for the onset of shear thinning, and is absolutely critical for understanding why deformation reduces nonexponential relaxation and narrows the distribution of relaxation times at remarkably low levels of stress or strain. 
\par
We have presented detailed numerical results for multiple questions including: deformation induced enhancement of mobility under stress, alpha relaxation time (or viscosity) power law shear thinning in the steady state, non-vanishing of the activation barrier in the shear thinning or beyond yield point regime, exponential growth of dynamic yield stresses with packing fraction, reduced fragility under deformation, and massive narrowing of the relaxation time distribution under deformation. The theory was shown to produce flow curves consistent with the empirical HB model over 5 decades in shear rate. All these aspects were studied as a function of fluid packing fraction, shear rate, and stress. The results seem to be broadly consistent with experiments and simulations. An important point is that activated dynamics generally remains present even at yielding, consistent with confocal experiments \cite{p19,p26}. 
\par
Future work in progress is focused on two main topics. First, we aim to construct a full rheological constitutive equation description of startup continuous shear of dense glass forming hard sphere colloidal suspensions. This will allow us to address questions such as stress overshoots and their microscopic connection to deformation induced changes of local packing structure \cite{p49}. Second, the ECNLE theory based approach will be extended to treat dense gels and attractive glasses characterized by both caging and strong physical bonding \cite{p58,p59} where a primary question is how such materials yield, including the molecular origin of the so-called “double yielding” behavior of attractive glasses \cite{p60,p61}.

\begin{acknowledgments}
This work was supported by DOE-BES under Grant DE-FG02-07ER46471 administered through the Materials Research Laboratory at UIUC. 
\end{acknowledgments}

\appendix

\section{Crowding vs Mechanically Driven Liquification of a Glass}
The phenomenological concept of “effective temperature” or “effective density” is sometimes invoked to describe how the application of mechanical forces changes dynamics and rheology \cite{p62}. Here we briefly consider this idea by comparing in the context of our theory how stress driven ‘liquification’ of a hard sphere glass relates to melting it by lowering density. 
\par
To conceptually confront the two perspectives, a generalized order parameter that defines a ‘glass’ kinetically from a fixed total barrier (essentially fixed time scale) criterion is defined. The total barrier corresponding to a high packing fraction of 0.61 is adopted as the criterion for the glass transition. For density driven liquification one must lower the packing fraction. This motivates defining $\theta_{\phi}=(\phi_g-\phi)/(\phi_g-\phi_{NMCT})$ where $\phi_g$ denotes the adopted vitrification criterion and $\phi_{NMCT}$ is the onset of activated dynamics. Thus, as  $\theta_\phi=0\rightarrow 1$, the total barrier decreases continuously from $\beta F_{Total}(\phi=0.61)\sim 27$ to 0. For the same packing fraction, an order parameter for stress driven change of dynamics is defined as $\theta_\Sigma=\Sigma/\Sigma_y^{abs}$ by noting that at $\Sigma=\Sigma_y^{abs}$ the barrier vanishes. Barriers in both cases are normalized, $f_{\Sigma}=F(\Sigma)/F(\Sigma=0)$ for stress driven changes, and $f_\phi=F(\phi)/F(\phi_g)$ for density driven changes. Study of such normalized barriers as a function of order parameter provides insight to the question of whether increasing stress mimics decreasing density.
\begin{figure}
    \centering
    \includegraphics[scale=0.76]{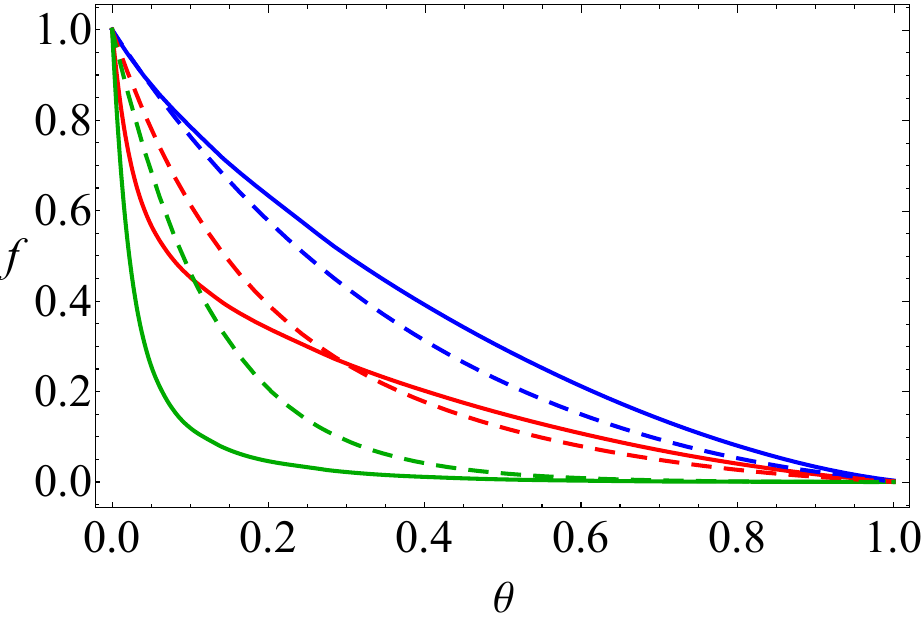}
    \caption{Doubly normalized plots for local (blue), elastic (green) and total (red) barriers as a function of ‘generalized order parameter’. The solid and dashed curves show ‘mechanically driven’ (stress) and ‘density driven’ (increasing crowding or, cooling) pathways to liquify a glass.}
    \label{figA1}
\end{figure}
\par
Figure 14 shows numerical results for the barriers using $\phi_g=0.61$. We do not find a strict collapse of the order parameters associated with the local, elastic or total barriers. This is especially true for the more accurate ECNLE theory that includes the collective elastic barrier. Hence, in the context of our theoretical approach we conclude that density driven changes of the activated relaxation process cannot serve as a reliably surrogate for barrier reduction via application of external stress.
\section*{References}
\bibliographystyle{apsrev4-1}
\bibliography{aipsamp}

\end{document}